\documentclass[journal]{IEEEtran}
\usepackage{epsfig}
\usepackage{graphicx}
\usepackage{color}
\usepackage{subfigure}

\usepackage[para]{threeparttable}
\usepackage{amssymb}
\usepackage{amsmath}
\usepackage{graphicx}
\usepackage{epsf}
\usepackage{epsfig}
\usepackage{psfig}
\usepackage{ccaption}
\usepackage{array}
\usepackage{tabularx}
\usepackage{multirow}
\usepackage{epsfig}
\usepackage{cite}
\usepackage{procedure,algorithm,algorithmic}

\ifCLASSINFOpdf

\else

\fi

\begin{document}
\twocolumn

\title{Time Synchronization Attack in Smart Grid-\\Part I: Impact and Analysis}

\author{Zhenghao Zhang,
        Shuping Gong,
        Aleksandar D. Dimitrovski,
        and Husheng Li,
\thanks{Z. Zhang, S. Gong and H. Li are with Department
of Electrical Engineering and Computer Science, University of Tennessee, Knoxville, TN.
A. D. Dimitrovski is with Energy and Transportation Sciences Division,
Oak Ridge National Lab, Oak Ridge, TN 37831
The research is under the support of National Science Foundation under
grants  ECCS-0901425.
}}

\markboth{{\em This paper has been submitted to IEEE Transaction on Smart Grid}}%
{Zhang \MakeLowercase{\textit{et al.}}: IEEE Transaction on Smart Grid}

\maketitle

\begin{abstract}
%\boldmath
Many operations in power grids, such as fault detection and event location estimation, depend on precise timing information. In this paper, a novel Time Synchronization Attack (TSA) is proposed to attack the timing information in smart grid. Since many applications in smart grid utilize synchronous measurements and most of the measurement devices are equipped with global positioning system (GPS) for precise timing, it is highly probable to attack the measurement system by spoofing the GPS. The effectiveness of TSA is demonstrated for three applications of phasor measurement unit (PMU) in smart grid, namely transmission line fault detection, voltage stability monitoring and event locationing. The validity of TSA is demonstrated by numerical simulations.
\end{abstract}


\IEEEpeerreviewmaketitle

\section{Introduction}
The research interest in smart grid \cite{Grid_future} has been growing in recent years.
As one of the key components in smart grid,
wide area monitoring systems (WAMSs) \cite{FNET_SGTRAN} have received tremendous attention.
The reliability of the smart grid system relies on the operation of WAMSs,
since the operations of smart grid demand the real-time
status of system provided by WAMSs.

WAMSs are typically constructed in a centralized manner.
The monitoring devices are placed throughout the entire smart grid system,
and they convey their measurement data to the control center
by certain communication infrastructure,
such as wireless network and optical fiber network.
The control center implements the analysis on these measurement data,
and corresponding control decisions will be made to maintain the normal operation of smart grid.
Note that supervisory control and data acquisition (SCADA) systems \cite{SCADA} have been
applied for maintaining the reliability of the power grid control systems.
However, SCADA mostly deals with random failures in the system, instead of malicious attacks.

The security of WAMSs is one of the key issues in smart grid technology,
since errors of monitoring measurements introduced by malicious attackers
will cause wrong control decisions, which may lead to a catastrophe like blackout \cite{BlackOut}.
\cite{DOS_attack} proposed a security strategy against denial-of-service (DoS) attack which focuses on
the cyber security of the communication infrastrcuture.
Meanwhile, malicious attack against measurement data, namely false data injection attack (FDIA)
has been studied in \cite{DataAttack_Kosut}\cite{DataAttack_LiuYao}\cite{DataAttack_Xie}.
By launching FDIA, malicious attackers can manipulate the system state variables
by modifying the measurements at a set of selected monitoring devices.
FDIA can mislead the control center to have an incorrect evaluation on the system operation status;
consequently wrong control decisions will be made.

To launch FDIA successfully, malicious attackers need to have
full knowledge of the power gird network such that a systematic false
measurements can be generated to bypass the bad measurement detection \cite{BadDataDetection_Lin}.
However, it is very difficult for attackers to obtain the full knowledge of the power grid network infrastructure
which can only be accessed by the power system operator.
In addition, FDIA requires physical accesses to several selected monitoring devices in order to inject the false measurement data.
This is another difficulty in practice, since those monitoring devices are typically placed at locations with physical security protection.

\begin{figure}[]
  \centering
  \includegraphics[scale=0.5]{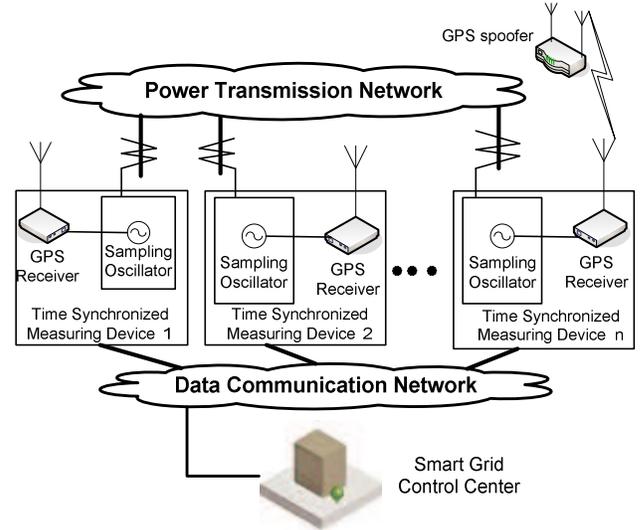}
  \caption{Illustration of time synchronized monitoring in smart grid with GPS spoofer}\label{Fig:SynMeasure}
\end{figure}

In this paper, we identify a potential attack to WAMSs in smart Grid, coined {\it time synchronization attack (TSA)}.
Note that monitoring devices are distributed throughout the entire power grid network,
whose measurements data are fed back to the control center with various transmission delays.
To obtain an accurate system operation status,
the control center needs to align all collected measurements in the time domain,
which is called time synchronized monitoring \cite{ImportantOfTimeSyn}.
Since global positioning system (GPS) signal is highly accurate and stable for timing without any extra communication infrastructure,
GPS based time synchronization monitoring devices have been vastly deployed in smart grid monitoring system.
Figure \ref{Fig:SynMeasure} illustrates time synchronized monitoring in smart grid.
There are $n$ time synchronized measuring devices (TSMD) installed throughout the entire smart grid system,
and each of them is equipped with a GPS signal receiver.
Note that TSMD is a general conception, which could be any measurement devices requiring time synchronization,
e.g. phase measurement units (PMU).
The grid operation state parameters, such as frequency and voltage, are sampled periodically
and the sampling is triggered by the GPS timing signal from the GPS receiver.
To cope with the different data transmission delays of different measurements,
it is necessary to attach the time values at which the measurements are sampled.
This procedure is similar to posting a stamp to the measurements (hence called time stamp).
The control center aligns the collected measurements according to their time stamps,
and analyzes the system state for future control actions.

By applying GPS timing as the grid-wide sampling reference time,
all TSMDs in the smart grid sample the observations in a synchronous manner.
However, a malicious attacker can modify the sampling time by introducing a forged GPS signal \cite{Spoofer}.
There are several studies that have identified the possibility of spoofing GPS receivers \cite{Spoofer}\cite{GPSSpoof_Motella}\cite{GPSSpoof_Tippenhauer}.
Furthermore, a realworld GPS spoofing attack was reported recently \cite{GPSSpoof_Goodin},
which demonstrated the vulnerability of GPS signals.
Note that the malicious attacker does not need to hack into the monitoring system or have physical contact to the TSMDs.
In addition, it is difficult to locate the malicious attacker since it can transmit the GPS spoofing signal as it moves around the target TSMD.
As illustrated in Figure \ref{Fig:SynMeasure},
the malicious attacker launches a TSA to one of the TSMDs by
transmitting counterfeit GPS signal, in which the timing has been modified.
The target TSMD will do sampling at a wrong time. Consequently, the measurements with false time stamps are conveyed to the control center.
The control center will therefore  misalign the measurements
and will obtain an incorrect system state.
Although there is some data processing procedure to handle the measurements,
most current processing schemes only consider the measurement error caused by noise
or packet loss; therefore, TSA can easily bypass a simple countermeasure scheme such as smoothing filtering.

Motivated by the security requirement of smart grid, in this paper, the impacts of TSA will be identified and the severeness of TSA will also be analyzed.
Specifically, we study TSA in three applications of PMU, namely transmission line fault detection/locationing, voltage stability monitoring and event locationing.
Moreover, TSA is not constrained to only PMU applications.
There exist potential TSA opportunities in any monitoring system requiring time synchronization.
Simulation results will demonstrate that TSA can effectively deteriorate the performance of these applications and may even result in false operations in power system.

The remainder of this paper is organized as follows.
Section \ref{sec: Attack model} provides the GPS spoofing attack model.
Section \ref{sec: LineFaultDet} studies the impacts of TSA on transmission line fault detection and fault localization.
The TSA damage analysis and corresponding simulation result of the voltage monitoring algorithm are presented in Section \ref{sec:VoltageStable}.
And Section \ref{sec:Event} presents the study of TSA in the task of
regional perturbation event location.
Conclusions and future work are provided in Section \ref{sec:Conclusion}.

\section{GPS Signal Receiving And Attack Model}\label{sec: Attack model}
In this section, we briefly introduce the GPS signal reception processing. Then we propose the attack model for GPS spoofing and TSA.

\begin{figure}[htp]
  \centering
  \includegraphics[scale=0.5]{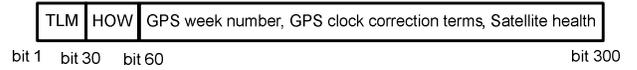}
  \caption{Subframe-1 structure}\label{Fig:Subframe}
\end{figure}
\subsection{Introduction of GPS Signal Receiving}
\begin{figure}[htp]
  \centering
  \includegraphics[scale=0.55]{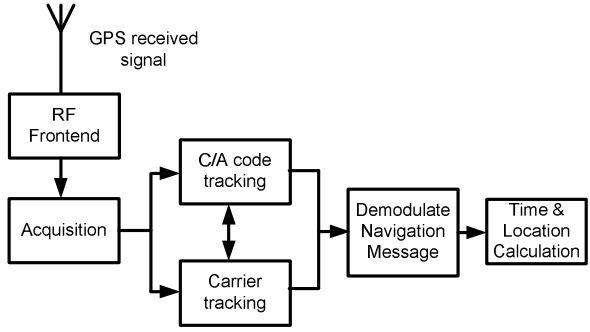}
  \caption{Diagram of GPS signal receiving processing}\label{Fig:GPS_recv}
\end{figure}

The precise timing information from GPS signals includes two parts: one is embedded in the navigation messages demodulated from the received GPS signals, whose precision is in the order of seconds; the other part is the precise signal propagation time from the GPS satellite to the receiver, which has the precision of millisecond for civilian users.
The timing information with precision of second is located in subframe $1$, whose frame structure \cite{Book_GPSreceiver} is illustrated in Figure \ref{Fig:Subframe},
where ``TLM'' is the telemetry data severing as preamble, and ``HOW'' provides the GPS time-of-week (TOW) modulo 6 seconds corresponding to the leading edge of the following subframe. Therefore, with TOW and GPS week number, we can obtain the date and the time with the precision of second. To obtain a more precise time value, we need to calculate the propagation time of the GPS signal from the satellite to the GPS receiver. Therefore, users in different locations can be synchronized by exploiting the GPS precise timing information as a time reference. The system-wide synchronization time reference is referred to the coordinated universal time (UTC) $t_{UTC}$ disseminated by GPS, which is given by
\begin{equation}\label{Eq. UTC}
    t_{UTC} = t_{rcv}-t_{p}-\Delta t_{UTC}.
\end{equation}
where $t_{rcv}$ and $t_{p}$ denote the receiver clock time and propagation time for the GPS signal, respectively;
and $\Delta t_{UTC}$ denotes the time corrections provided by the GPS ground controllers.
To obtain the navigation message, we need to demodulate the GPS signal.
A typical GPS signal reception processing is illustrated in Figure \ref{Fig:GPS_recv}.

The received standard positioning service (SPS) GPS signal $r(t)$ is given by
\begin{equation}\label{Eq. GPS_signal}
    r(t) = \sum_{k=1}^{32}H_k(2P_c)^{1\over 2}(C_k(t)\oplus D_k(t))cos2\pi (f_{L1}+\Delta f_k)t+n(t),
\end{equation}
where $H_k$ and $P_c$ are the channel matrix for the $k$-th satellite and the signal power, respectively;
$C_k(t)$ and $D_k(t)$ are the spread spectrum sequence (C/A code) and the navigation message data from the $k$-th satellite, respectively;
$f_{L1}$ and $\Delta f_k$ are the carrier frequency for civilian GPS signal and doppler frequency shift for the $k$-th satellite, respectively; and $n(t)$ is noise.
As illustrated in Figure \ref{Fig:GPS_recv}, the signal processing includes two major steps, namely acquisition and tracking.
From (\ref{Eq. GPS_signal}), we can observe that the key processing for acquisition is to search for the code phase of the received
C/A code and doppler frequency shift $\delta f_k$. By multiplying the C/A code of identical code phase and the carrier of identical frequency with the received GPS signal, the navigation message can be demodulated coherently \cite{Book_GPSreceiver}.

\subsection{Attack model}
\begin{figure}[]
\vspace{0pt}
\subfigure[No attack\label{Corr_ture}]{
%\begin{minipage}[tpb]{11cm}
\begin{minipage}[b]{0.5\linewidth}
\centering
\includegraphics[scale=0.3]{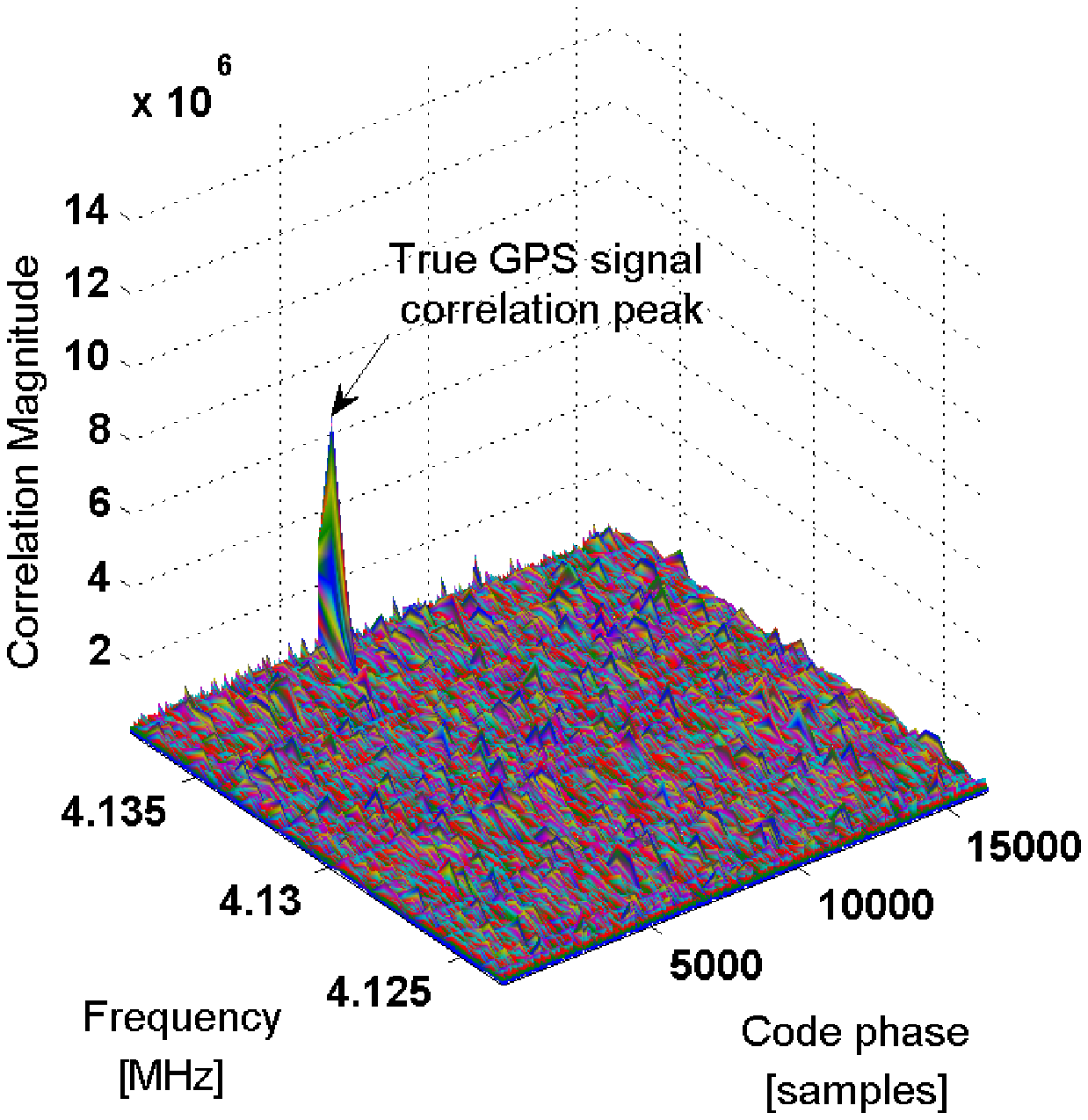}
\end{minipage}}%
%\hfill \vspace{0pt}
\subfigure[Under spoofing attck\label{Corr_attack}]{
%\begin{minipage}[tpb]{11cm}
\begin{minipage}[b]{0.5\linewidth}
\centering
\includegraphics[scale=0.28]{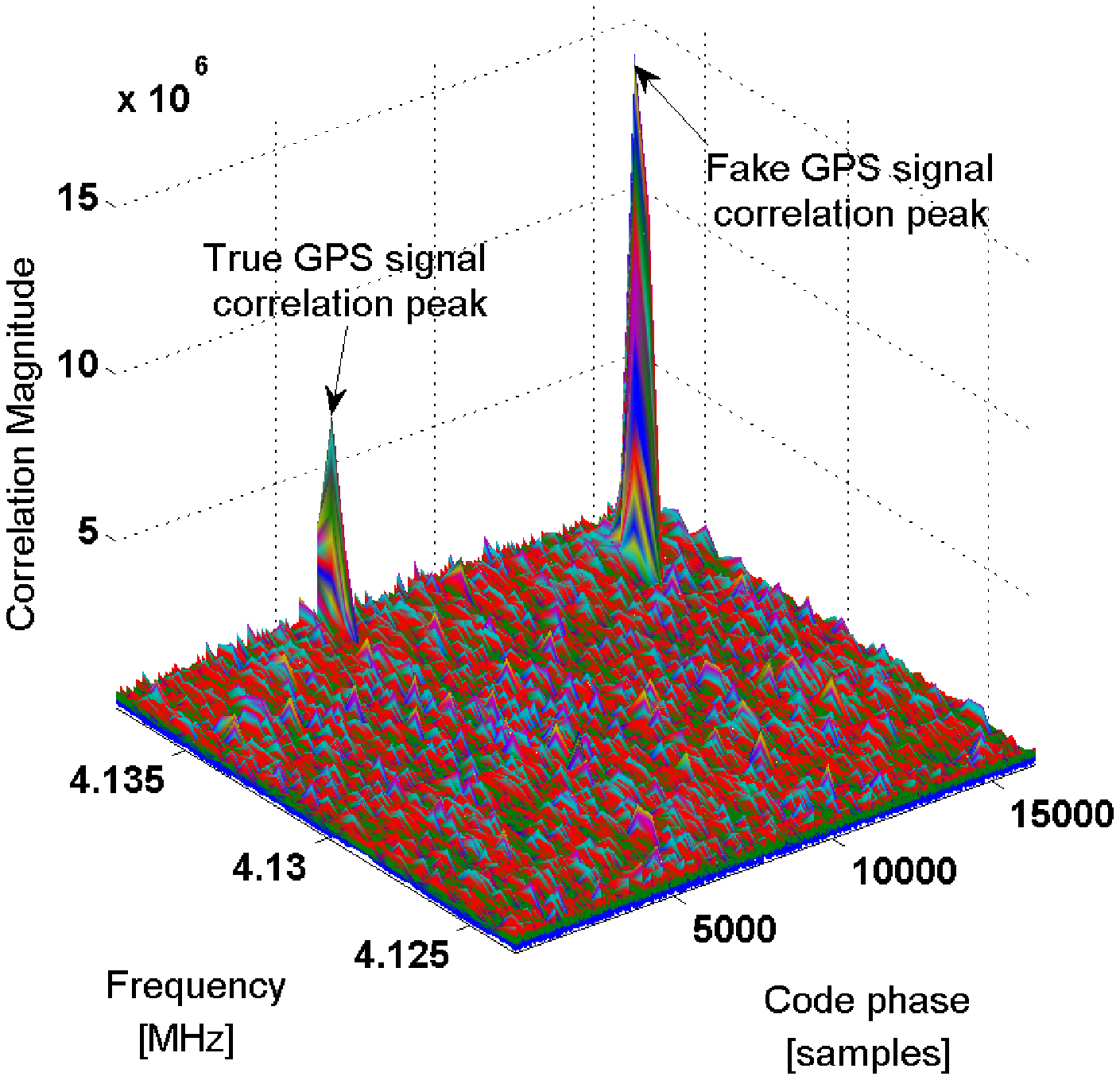}
\end{minipage}}
\caption{Comparison of the correlation peak under normal and spoofing attack reception conditions.} \label{Fig_CorrPeak}
\end{figure}

To spoof a GPS receiver, the GPS receiver needs to be misled to acquire the fake GPS signal instead of the true one.
Since the acquisition is implemented by searching for the highest correlation peak in the code-phase-carrier-frequency two dimensional space,
intuitively, the signal with higher signal-to-noise-ratio (SNR) will have a higher correlation peak, which is illustrated in Figure \ref{Fig_CorrPeak}.
Therefore, there exists a two-step spoofing strategy. In the first step, the spoofer launches certain interference which causes the GPS receiver to lose track. In the second step, it launches the spoofing GPS signal when the GPS receiver carries out the acquisition processing. Consequently, the GPS receiver will track the counterfeit GPS signal due to its higher correlation peak, since the counterfeit GPS signal has a higher SNR.

\begin{figure*}[]
\vspace{0pt}
\subfigure[First stage: correlation peak scanning\label{hooking}]{
%\begin{minipage}[tpb]{5cm}
\begin{minipage}[b]{0.3\linewidth}
\centering
\includegraphics[scale=0.5]{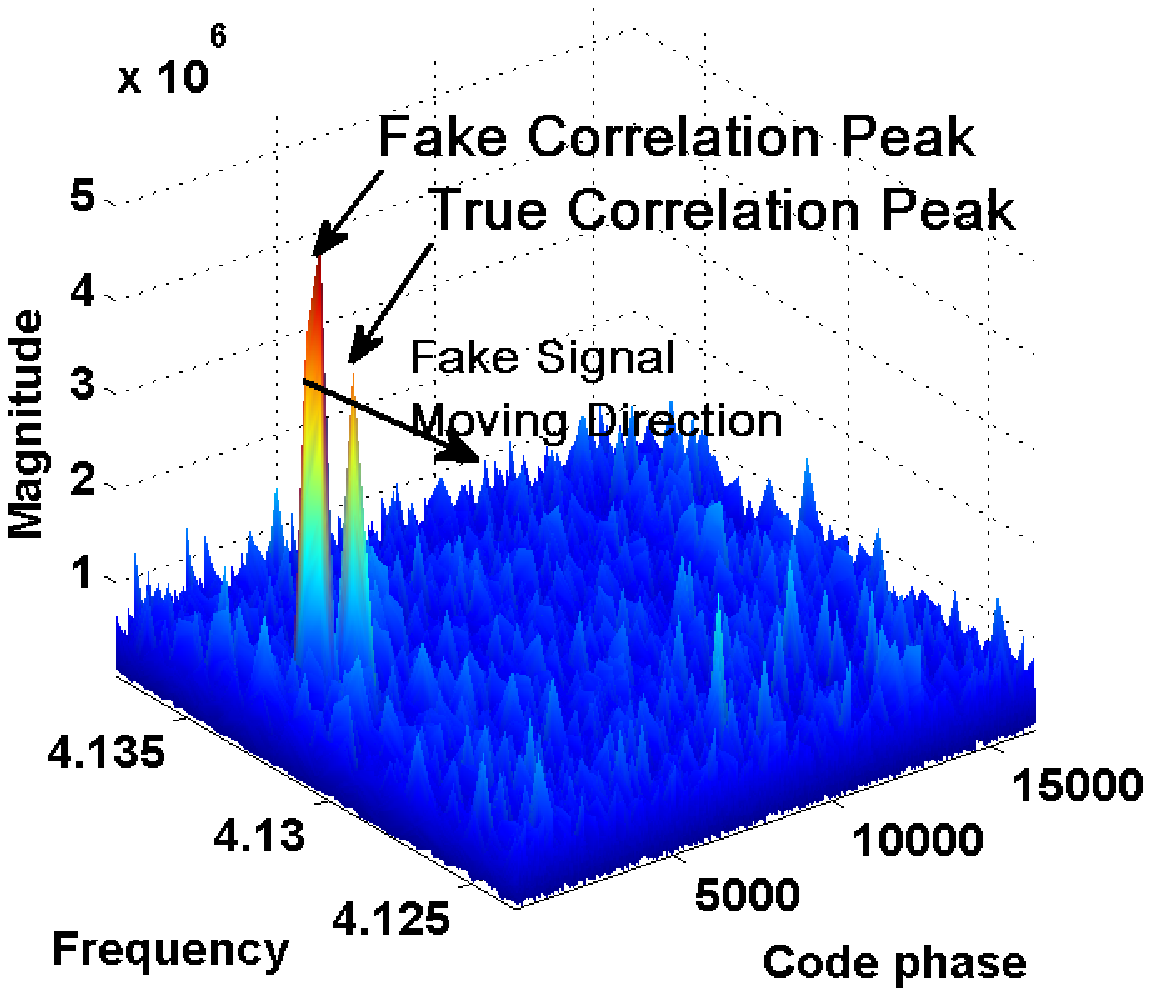}
\end{minipage}}%
%\hfill \vspace{0pt}
\hspace{0.5cm}
\subfigure[Second stage: fake correlation peak overlaps the fake one\label{overlap}]{
%\begin{minipage}[tpb]{11cm}
\begin{minipage}[b]{0.3\linewidth}
\centering
\includegraphics[scale=0.5]{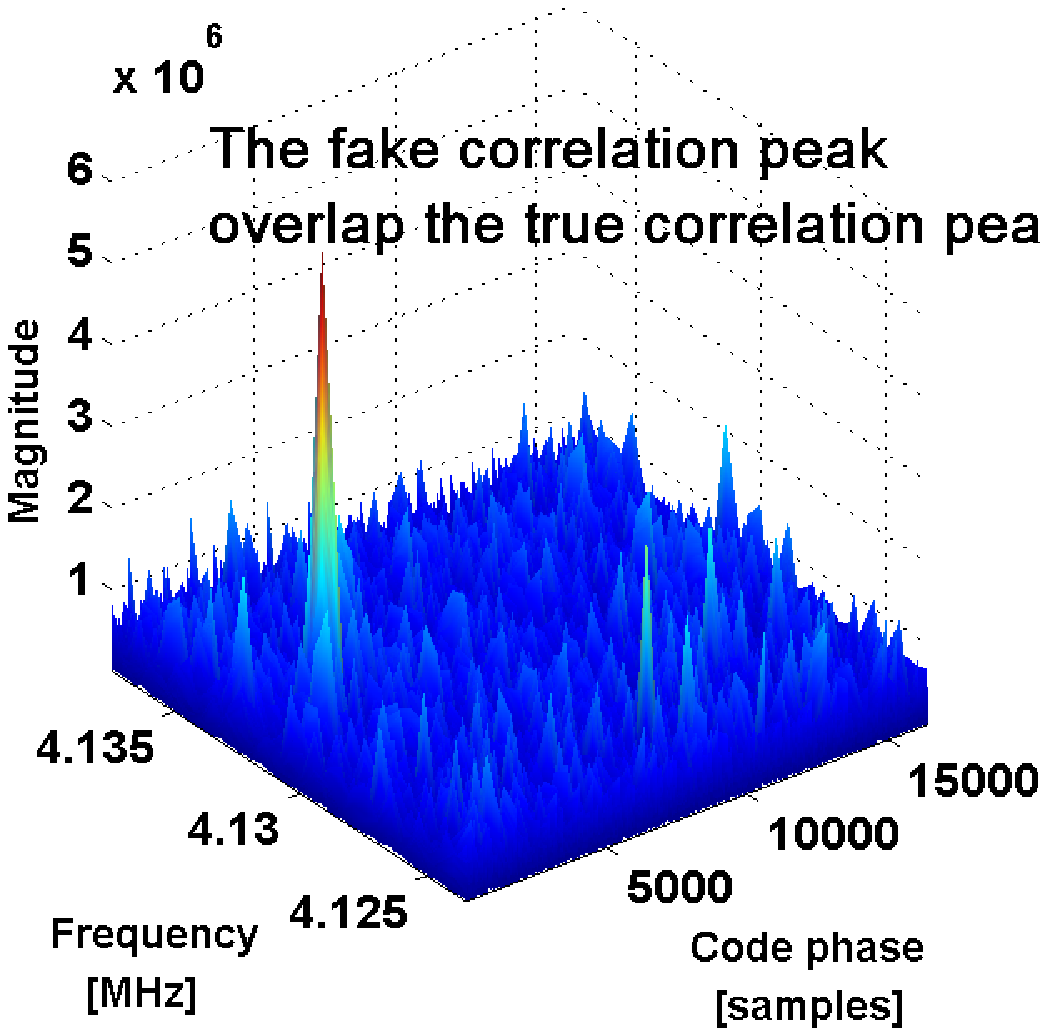}
\end{minipage}}
\hspace{0.5cm}
\subfigure[Third stage: move the fake correlation peak to the attacker's desired point\label{captured}]{
%\begin{minipage}[tpb]{11cm}
\begin{minipage}[b]{0.3\linewidth}
\centering
\includegraphics[scale=0.5]{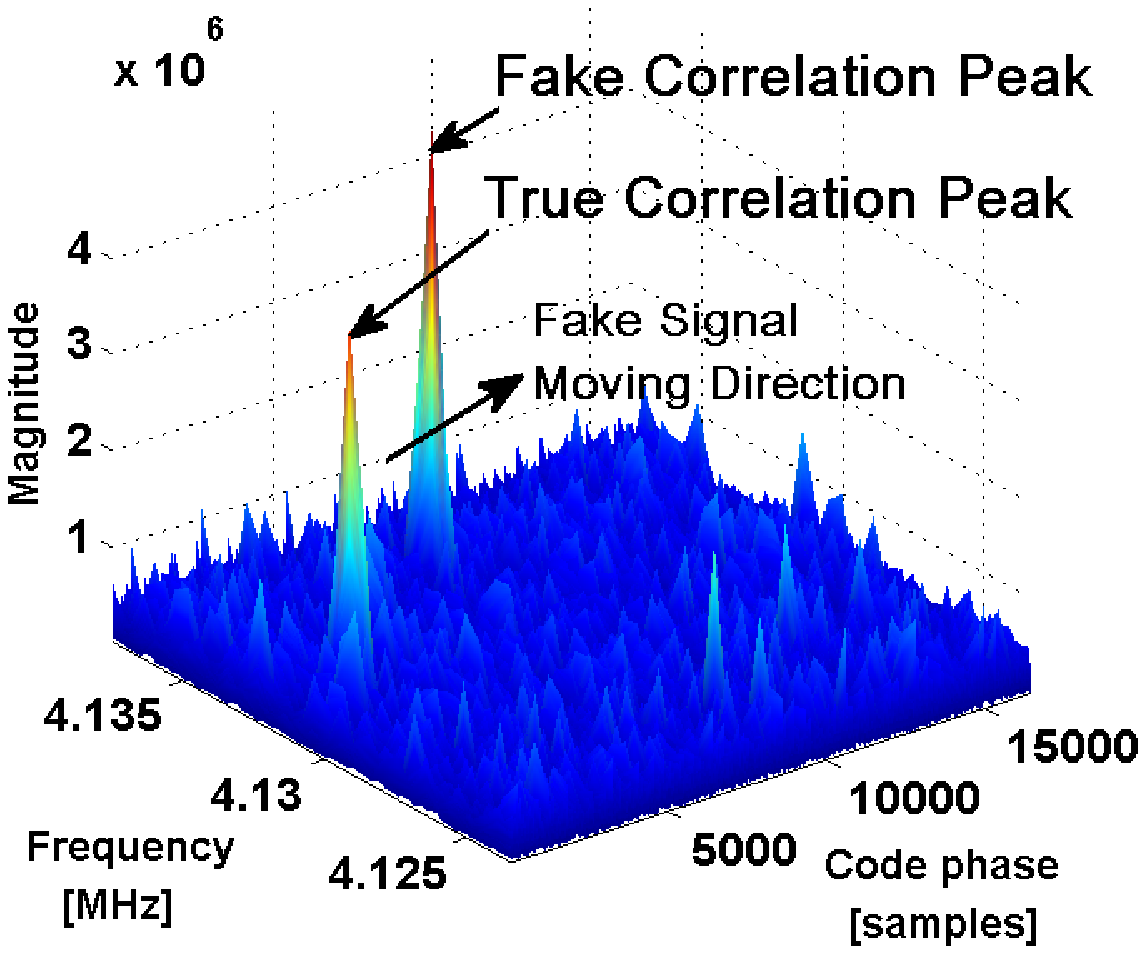}
\end{minipage}}
\caption{Spoof the GPS receiver by a three-stage attack.} \label{Fig_ScanAttck}
\end{figure*}

Alternatively, the attacker can scan the two-dimensional space of code phase and carrier frequency until the fake correlation peak overlaps the true correlation peak, which is illustrated in Figure \ref{Fig_ScanAttck}.
The first stage is correlation peak scanning, in which attacker launches the fake correlation peak close to the true correlation peak and moves slowly towards the true correlation peak.
Note that it is not difficult for the malicious attacker to estimate the location of the target GPS receiver, such that it can obtain the information of the true correlation peak by inducing from its own GPS receiver. Therefore, the attacker does not need to implement blind search on the entire two-dimensional space of code phase and carrier frequency.
In the second stage, the fake correlation peak moves to the position in which the fake correlation peak overlaps the true one.
The GPS receiver will be captured by the counterfeit signal and locked to the fake correlation peak, since it has a higher SNR.
In the third stage, the attacker will move the fake correlation peak slowly to the desired point.
At this time, the true correlation peak will be considered as noise.

\section{TSA in Transmission Line Fault detection and Fault Localization}\label{sec: LineFaultDet}
In this section, we study the impact of TSA on transmission line fault detection and localization.
Since a fault of a single transmission line may trigger cascading failures spreading within the entire power grid system,
it requires quick and accurate locationing of the fault in a wide power grid area.
One conventional method is to detect and localize the fault by utilizing local voltage and current measurements.
For improving the accuracy and locationing speed, many researchers proposed to utilize measurements at both ends of transmission line \cite{Novosel1996}\cite{Jiang2000}\cite{Liao2007}.
These measurements are attached with sampling time which is obtained from its GPS signal receiver; therefore TSA can affect the fault detection and localization of transmission lines.
In this section, we will first briefly review the fault detection and location in transmission line.
Then, the impact of TSA on the transmission line fault detection and location will be analyzed.
Simulations results will be provided at the end of this section.

\subsection{Fault Detection and Fault Localization for Long Transmission Line}\label{subsec:linefault}
\begin{figure}
  \centering
  \includegraphics[scale=0.6]{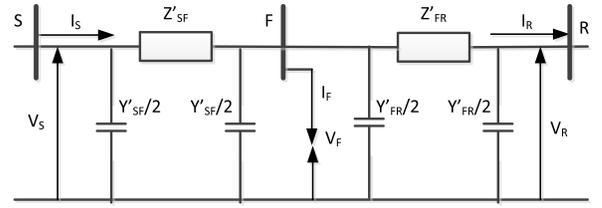}
  \caption{Model for long transmission line model with fault}\label{fig:m_fault_l}
\end{figure}

The model of long transmission line with fault \cite{Abboud1}\cite{Jiang2000} is shown in Figure \ref{fig:m_fault_l}.
Suppose that the total length of transmission line is $L$, and $F$ is the fault location.
As is shown in Figure \ref{fig:m_fault_l},
the fault point $F$ divides the whole transmission line into two sections,
which include line section $SF$ and line section $FR$.
The transmission line sections $SF$ and $FR$ can still be considered as two perfect transmission lines.
We define the fault location index $D\in[0,1]$ such that the distance from the fault location $F$ to the receiving end $R$ is $DL$.
On both sides of the fault point, the transmission line is represented by an equivalent $\pi$ circuit \cite{Book_PowerSystem}.
On the transmission line section $SF$, the sending end voltage of the equivalent $\pi$ circuit $V_S$ is given by
\begin{equation}\label{Eq.V_SF}
    V_S=\left(\frac{Z_{SF}^\prime Y_{SF}^\prime}{2}+1\right)V_F+Z_{SF}^\prime I_S,
\end{equation}
where $V_F$ and $I_F$ are the voltage and the current at the fault location, respectively;
$Z_{SF}$ and $Y_{SF}$ are the equivalent series arm impedances and the equivalent shunt arms
admittances of transmission line section $SF$, respectively.
Similarly, in the transmission line section $FR$, the sending end voltage of the equivalent $\pi$ circuit $V_F$ is given by
\begin{equation}\label{Eq.V_FR}
    V_F=\left(\frac{Z_{FR}^\prime Y_{FR}^\prime}{2}+1\right)V_R+Z_{FR}^\prime I_R,
\end{equation}
where $V_R$ and $I_R$ are the voltage and the current at the receiving end of the transmission line, respectively;
$Z_{FR}$ and $Y_{FR}$ are the equivalent series arm impedances and the equivalent shunt arms
admittances of transmission line section $FR$, respectively.
The equivalent series arm impedances $Z_{SF}$
and $Z_{FR}$ are given by
\begin{eqnarray}\label{Eq.Z_FR_SF}
  Z_{SF}^\prime &=& Z_{SF} \frac{\sinh(\gamma (1-D)L)}{\gamma (1-D)L} \\
  Z_{FR}^\prime &=& Z_{FR} \frac{\sinh(\gamma DL)}{\gamma DL}
\end{eqnarray}
with
\begin{eqnarray}
  Z_{SF} &=& (1-D)L\bar{z} \\
  Z_{FR} &=& DL\bar{z} \\
  \gamma &=& \sqrt{\bar{z}\bar{y}}
\end{eqnarray}
where $Z_{SF}$ and $Z_{FR}$ are the total series impedance of the line sections $SF$
and $FR$, respectively; $\bar{z}$ and $\bar{y}$ are the unit line impedance and admittance, respectively; and $\gamma$ is called the attenuation constant.
The equivalent shunt arms admittance $Y_{SF}$
and $Y_{FR}$ are given by
\begin{eqnarray}
Y_{SF}^\prime &=& Y_{SF} \frac{\tanh(\frac{\gamma (1-D)L}{2})}{\frac{\gamma (1-D)L}{2}}\\
Y_{FR}^\prime &=& Y_{FR} \frac{\tanh(\frac{\gamma DL}{2})}{\frac{\gamma DL}{2}}
\end{eqnarray}
with
\begin{eqnarray}
  Y_{SF} &=& (1-D)L\bar{y} \\
  Y_{FR} &=& DL\bar{y}
\end{eqnarray}
where $Y_{SF}$ and $Y_{FR}$ are the shunt arms admittance of the line section $SF$
and $FR$, respectively.

When fault occurs, the voltages $V_F$ at the fault location calculated from
(\ref{Eq.V_SF}) and (\ref{Eq.V_FR}) are identical \cite{Jiang2000}. Thus, substituting (\ref{Eq.V_FR}) into (\ref{Eq.V_SF}), the fault location index $D$ can be estimated as
\begin{eqnarray}
D_e = \frac{\ln(N/M)}{2\gamma L}
\end{eqnarray}
where
\begin{eqnarray}
M &=& \frac{V_S + Z_c I_S}{2}\exp(-\gamma L) - \frac{V_R + Z_c I_R}{2}\label{Eq.M}\\
N &=& \frac{V_R - Z_c I_R}{2} - \frac{V_S - Z_c I_S}{2}\exp(\gamma L)\label{Eq.N}
\end{eqnarray}
where $Z_c = \sqrt{z_1/y_1}$ is the characteristic impedance of transmission line.
Furthermore, it can be observed from (\ref{Eq.M}) and (\ref{Eq.N}) that,
if there is no fault, the computed absolute values of $M$ and $N$ are all held at zero.
Therefore, $M$ and $N$ can also be utilized as fault indicators \cite{Jiang2000}.

In practice, PMUs are installed at both ends of the transmission line to obtain
$V_S$, $V_R$, $I_S$, and $I_R$. These measurements will be conveyed to the control center along with their time stamps.
Control center will exploit the time stamps of these measurements for alignment such that the indicators $N$ and $M$
can be calculated in terms of the measurements sampled from at the same time. In the next subsection, we will analyze how TSA affects the transmission line fault detection and fault location.

\subsection{Analysis of Impact}
In this subsection, we analyze the impact of TSA on the transmission line fault detection and location. The transmission line fault detection and location is based on the PMUs installed on both ends of the transmission line.
It should be noted that the measurements $V_S$, $V_R$, $I_S$, and $I_R$ have complex values.
When TSA is launched toward target PMUs, the time stamps on these measurements will be modified,
which is equivalent to modifying the phase angle of these measurements. The phase angle errors resulted from TSA at the sending PMU and receiving PMU are denoted by $\Delta\theta_S$ and $\Delta\theta_R$, respectively. And the measurements $V_S$, $V_R$, $I_S$, and $I_R$ affected by TSA are denoted as $\tilde{V_S}$, $\tilde{V_R}$, $\tilde{I_S}$, and $\tilde{I_R}$,
which are given by
\begin{eqnarray}
  \tilde{V_S} &=& |V_S|\exp j(\theta_{V_S}+\Delta\theta_S)\nonumber\\
              &=& V_S\exp(j\Delta\theta_S)\label{Eq.TildeVS}\\
  \tilde{V_R} &=& |V_R|\exp j(\theta_{V_R}+\Delta\theta_R)\nonumber\\
              &=& V_R\exp(j\Delta\theta_R)\label{Eq.TildeVR}\\
  \tilde{I_S} &=& |I_S|\exp j(\theta_{I_S}+\Delta\theta_S)\nonumber\\
              &=& I_S\exp(j\Delta\theta_S)\label{Eq.TildeIS}\\
  \tilde{I_R} &=& |I_R|\exp j(\theta_{I_R}+\Delta\theta_R)\nonumber\\
              &=& I_R\exp(j\Delta\theta_R)\label{Eq.TildeIR}
\end{eqnarray}

To analyze the impact of TSA on the transmission line fault detection, we substitute (\ref{Eq.TildeVS})-(\ref{Eq.TildeIR}) into (\ref{Eq.M}) and (\ref{Eq.N}) and then obtain
\begin{eqnarray}
  \tilde{M} &=& \frac{V_S + Z_c I_S}{2}\exp(-\gamma L)\exp(j\Delta\theta_S)\nonumber\\
            & & - \frac{V_R + Z_c I_R}{2}\exp(j\Delta\theta_R)\label{Eq.TildeM} \\
  \tilde{N} &=& \frac{V_R - Z_c I_R}{2}\exp(j\Delta\theta_R) \nonumber\\
            & & - \frac{V_S - Z_c I_S}{2}\exp(\gamma L)\exp(j\Delta\theta_S).\label{Eq.TildeN}
\end{eqnarray}

The impacts of TSA on the line fault detection indicators $M$ and $N$ are equivalent
to adding amplitude modulations. The error of line fault location due to TSA is
given by

\begin{eqnarray}
  \Delta D &=& D-D_{TSA} \nonumber\\
    &=& ({1\over 2\gamma L})
        \ln(\frac{N}{M}\frac{\tilde{M}}{\tilde{N}})\nonumber\\
    &=& ({1\over 2\gamma L})
        \ln(\frac{(A+B)(C+D\epsilon)}{(C+D)(A+B\epsilon)})
\end{eqnarray}

with
\begin{eqnarray}
  A &=& V_R - Z_c I_R\\
  B &=& - (V_S - Z_c I_S)\exp(\gamma L) \\
  C &=& - (V_R + Z_c I_R) \\
  D &=& (V_S + Z_c I_S)\exp(-\gamma L)\\
  \epsilon &=& \exp(j(\Delta\theta_R-\Delta\theta_S))=\exp(j\Delta\theta),
\end{eqnarray}
where $\Delta\theta$ denotes the asynchronisim of the phase angles of the measurements between the sending end and the receiving end caused by TSA. In the next subsection, the simulation results will show that the attacker can obtain the maximum line fault location error by launching TSA jointly on both the sending and receiving ends simultaneously.

%-------------------------------------------------------------------------------------------------------------------------

\subsection{Simulation results of TSA on transmission line fault detection and location }
\begin{figure}
  \centering
  \includegraphics[scale=0.75]{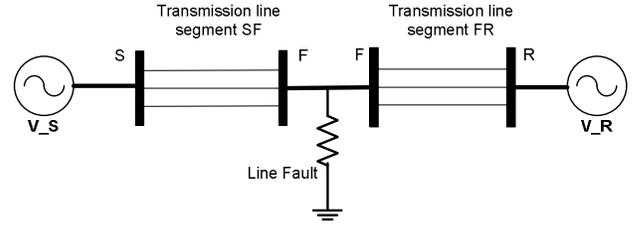}
  \caption{Simulation model for transmission line fault detection and location}\label{fig:sim_f_model}
\end{figure}

In this section, simulations have been conducted to evaluate the impacts of TSA on the
transmission line fault detection and fault location.
The simulation model for transmission line is shown in Figure \ref{fig:sim_f_model}.
The parameters used for the transmission line are listed in Table \ref{Tab: LineSimSettings},
which are the same as those in \cite{Novosel1996}.

%% The table of the setting
%************************************ Tab 1**************************************************
\begin{table}[t]
\newcommand{\rb}[1]{\raisebox{1.5ex}[0pt]{#1}}
\centering
\caption{Simulation settings for transmission line fault detection and location}
\begin{tabular}{|c|c| }
\hline
                 {Parameters}&{Setting values}\\
\hline
                 {Sending End}    & \\
                 {Voltage $V_S$}  &  $25000$(V) \\
\hline
                 {Receiving End}   &\\
                 {Voltage $V_R$}   &  $20000$(V) \\
\hline
                {Frequency}   &  $60(Hz)$ \\
\hline
                {Transmission}  &\\
                {line length}   &  $400$(km) \\
\hline
                {Transmission}      &  \\
                {line resistance}   &  $0.249168+j0.60241(Ohms/km)$ \\
\hline
                {Transmission}      &\\
                {line inductance}   &  $0.00156277+j 0.60241(H/km)$ \\
\hline
                {Transmission}       &\\
                {line capacitance}   & $19.469\times10^{-9} + j 12.06678\times10^{-9}(F/km)$ \\
\hline
\end{tabular}\label{Tab: LineSimSettings}

\end{table}
%%%%%%%%%%%%%%%%%%%%%%%%%%%%%%%%%%%%%%%%%%%%%%%%%%%%%%%%%%%%%%%%%%%%%%%%%%%

\begin{figure}
  \centering
  \includegraphics[scale=0.42]{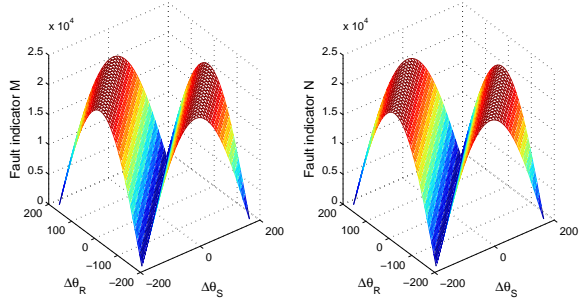}
  \caption{Impacts of TSA on transmission line fault indicator}\label{fig:TSA_FaultIndicator}
\end{figure}

Firstly, we study the impact on the fault indicator.
Figure \ref{fig:TSA_FaultIndicator} shows the TSA impacts on the
fault indicators $M$ and $N$ when various $\Delta\theta_S$ and $\Delta\theta_R$
are applied for TSA.
From (\ref{Eq.M}) and (\ref{Eq.N}), $M$ and $N$ should both hold on zeros,
when there is no transmission line fault.
However, when malicious attackers launch TSA cooperatively
on both the sending and receiving ends of the transmission line, the
attackers can modify the fault indicator value. Consequently,
TSA may trigger false alarm at the control center.

\begin{figure}
  \centering
  \includegraphics[scale=0.5]{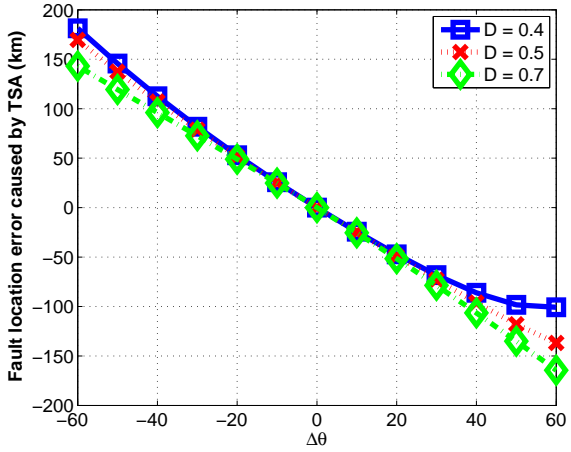}
  \caption{Impacts of TSA on transmission line fault location}\label{fig:D_error}
\end{figure}

We simulate the scenario when there is a three-phase grounded transmission line fault.
Figure \ref{fig:D_error} demonstrates the TSA impact on the transmission line fault location.
We simulate various scenarios in which the line fault occurs in different locations.
From Figure \ref{fig:D_error}, we observe that TSA can produce fault location error as large as 180km.
Notice that it is important to locate the fault accurately in a short time; otherwise,
the local line fault may lead to network-wide cascading fault.
Therefor the error caused by TSA will severely affect the system-wide reliability of smart grid.

\begin{figure}
  \centering
  \includegraphics[scale=0.5]{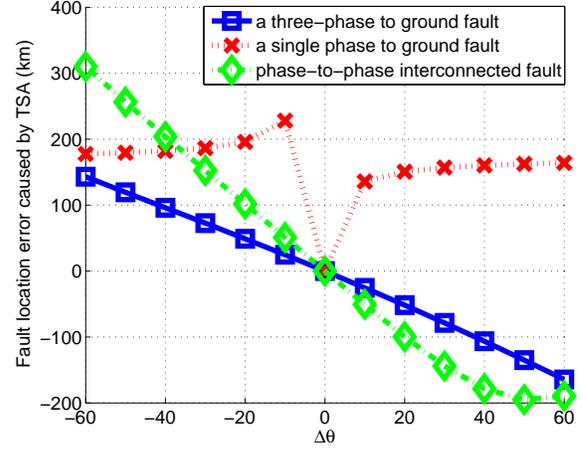}
  \caption{Impacts of TSA on various types of transmission line faults}\label{fig:VarFault}
\end{figure}

Figure \ref{fig:VarFault} demonstrates the TSA impacts on various types of transmission line faults.
It is observed from Figure \ref{fig:VarFault} that TSA has different impact patterns for
different types of transmission line faults. However, the malicious attacker can always launch a TSA
causing the maximal error to the transmission line fault location by cooperatively attacking the sending and receiving ends.

\section{TSA in Voltage Stability Monitoring}\label{sec:VoltageStable}
Voltage stability monitoring is one of the key tasks in smart grid.
One commonly used method to evaluate the voltage stability is to apply T-equivalent and
Thevenin equivalent circuit to
set up a simplified model for power system \cite{Larsson2002}. The key idea is to apply GPS based synchronized PMU to monitor
the voltage and current in order to obtain the voltage stability indicators.
In this section, we study the impact of TSA on
the voltage stability monitoring.

\subsection{Model of Voltage Stability Monitoring}
\begin{figure}
  \centering
  \includegraphics[scale=0.6]{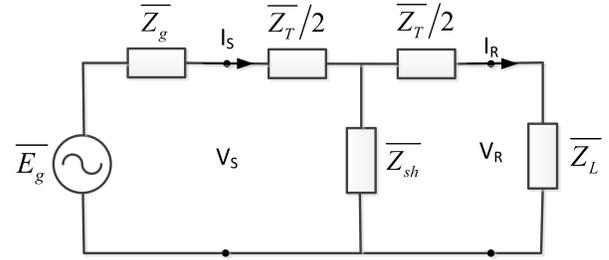}
  \caption{T-equivalent circuit for power system}\label{fig:vol_Teq}
\end{figure}

The simplified power system modeling for voltage stability monitoring includes
two key stages. The first stage is to calculate the parameters of
a T-equivalent of the actual transmission corridor with the GPS based
synchronized measurements \cite{Larsson2002}.
Figure {\ref{fig:vol_Teq}} illustrates the T-equivalent circuit.

In the T-equivalent circuit, the whole network is divided into three parts: generation source $\bar{E}_g$ with impedance $\bar{Z}_g$, transmission network and local load.
The available measurements include local measurements $V_R$, $I_R$, and remote measurements $V_S$, $I_S$ which are associated with the generation source and the transmission network.
These measurements can be sampled by PMU and be conveyed to the control center along with their time stamps.
The control center aligns these measurements according to their time stamps and obtains the system operation parameters
$\bar{Z}_T$, $\bar{Z}_{sh}$ and $\bar{Z}_{L}$, which can be estimated as follows:
\begin{eqnarray}
\bar{Z}_T &=& 2 \frac{V_S - V_R}{I_S + I_R} \\
\bar{Z}_{sh} &=& - \frac{V_S I_R + V_R I_S}{I_R^2 - I_S^2}\\
\bar{Z}_{L} &=& \frac{V_R}{I_R}.
\end{eqnarray}

The complex valued generator voltage $\bar{E}_g$ and its equivalent impedance $\bar{Z}_g$
cannot be estimated simultaneously. However, in practical cases, $\bar{Z}_g$ is assumed to be known by the characteristics of the step-up transformers and the transmission line.
Thus, the equivalent complex voltage of the generators is given by
\begin{eqnarray}\label{Eq.Eg}
\bar{E}_g = V_S + I_S \bar{Z}_g.
\end{eqnarray}

After calculating the parameters of the T-equivalent circuit, the Thevenin equivalent circuit is applied to further simplify the power system model. $\bar{E}_{th}$ and $\bar{Z}_{th}$ are associated with the following equation:
\begin{eqnarray}
\bar{E}_{th} = V_R + \bar{Z}_{th} I_R,
\end{eqnarray}
where $\bar{E}_{th}$ and $\bar{Z}_{th}$ are the equivalent voltage source and the equivalent source impedance in the Thevenin equivalent circuit, which can be calculated by the parameters of the T-equivalent circuit:
\begin{eqnarray}
\bar{E}_{th} &=& V_R \frac{\bar{Z}_{th} + \bar{Z}_L}{\bar{Z}_L}\\
\bar{Z}_{th} &=& \frac{\bar{Z}_T}{2} + \frac{1}{\frac{1}{Z_{sh}} + \frac{1}{Z_T/2+\bar{Z}_g}}.
\end{eqnarray}

When there are transmission lines tripped, the system voltage will become unstable.
If the malfunction is not repaired in time, the entire system will eventually collapse.
With the Thevenin equivalent circuit, two important stability margins can be obtained \cite{Larsson2002}.
The first indicator is associated with load impedance, which is given by
\begin{eqnarray}\label{Eq.MZ}
\text{MARGIN}_Z = 100 (1 - k_{\text{crit}}),
\end{eqnarray}
where
\begin{eqnarray}
k_{\text{crit}} = \left|\frac{\bar{Z}_{th}}{\bar{Z}_L}\right|.
\end{eqnarray}

Assuming that the type of load is constant power consumer,
we define a scale factor $k$ which is used to model the change in the load impedance.
We can set $\bar{Z}_L=k\bar{Z}_{L0}$, where $\bar{Z}_{L0}$ represents the value of
load impedance.
The transfer power is given by
\begin{equation}\label{Eq.PL}
    P_{L} = \Re\left(k\bar{Z}_{L0}\left|\frac{\bar{E}_{th}}{\bar{Z}_{th} + \bar{Z}_{L0}}\right|^2\right).
\end{equation}
Substituting $k=k_{crit}$ into (\ref{Eq.PL}), we can obtain the maximum possible power transfer, which is given by
\begin{equation}\label{Eq.PMax}
    P_{Lmax} = \Re\left(k_{crit}\bar{Z}_{L0}\left|\frac{\bar{E}_{th}}{\bar{Z}_{th} + k_{crit}\bar{Z}_{L0}}\right|^2\right)
\end{equation}

The second indicator is associated with the active power delivered to the load bus, which is given by
\begin{eqnarray}
\text{MARGIN}_P = \left\{
\begin{array}{ll}
p_{\text{Lmax}} - P_L, & \text{if } \bar{Z}_L > \bar{Z}_{th} \\
0, & \text{if } \bar{Z}_L > \bar{Z}_{th}
\end{array}
\right..
\end{eqnarray}

\subsection{Analysis of Impact}
TSA affects the time stamps of the monitoring measurements similarly
to the analysis in (\ref{Eq.TildeVS})-(\ref{Eq.TildeIR}).
It will modify the local and remote monitoring measurements by modifying their phase angles. It can be observed that all the voltage stability monitoring indicators are based on the T-equivalent parameters $\bar{Z}_T$, $\bar{Z}_{sh}$, and $\bar{Z}_L$.
Under TSA, these three parameters are modified to
\begin{eqnarray}
\bar{Z}'_T &=& 2 \frac{V_S \exp(j\Delta\theta_S) - V_R\exp(j\Delta\theta_R)}
                {I_S\exp(j\Delta\theta_S) + I_R\exp(j\Delta\theta_R)} \\
\bar{Z}'_{sh} &=& - \frac{(V_S I_R + V_R I_S)}{I_R^2\exp(j2\Delta\theta_R) - I_S^2\exp(j2\Delta\theta_S)}\nonumber\\
& &\times(\exp j(\Delta\theta_S+\Delta\theta_R))\\
\bar{Z}'_{L} &=& \frac{V_R\exp(j\Delta\theta_R)}{I_R\exp(j\Delta\theta_R)}=\bar{Z}_{L}.
\end{eqnarray}
It can be observed that the TSA affects both $\bar{Z}_T$ and $\bar{Z}_{sh}$.
Furthermore, it concerns the Thevenin equivalent circuit parameters $\bar{Z}_{th}$
and $\bar{E}_{th}$. Since $\bar{Z}_{th}$ depends on the calculation result of
the T-equivalent parameters $\bar{Z}_T$ and $\bar{Z}_{sh}$, the Thevenin equivalent
impedance will be substantially affected by TSA.
Consequently, TSA affects the entire calculation of the indicators of voltage stability monitoring. In the next subsection, simulation results will demonstrate the TSA impacts.

\subsection{Simulations of Voltage Stability Monitoring under TSA}
\begin{figure}
  \centering
  \includegraphics[scale=0.65]{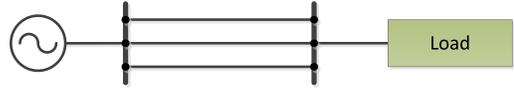}
  \caption{Simulation model for voltage stability}\label{fig:sim_v_model}
\end{figure}

The simulation model for voltage stability monitoring is shown in Fig. \ref{fig:sim_v_model}. The root mean square amplitude of source voltage dynamically changes with frequency 1Hz. The load has constant power comsuption. There are three transmission lines. A type phase ABC short-circuit fault occurs on transmission line 1 between 2 seconds and 2.5 seconds. Transmission lines 1 and 2 are tripped at time $4$ seconds and $6$ seconds.

\begin{figure}[]
\vspace{0pt}
\subfigure[TSA impact on $Z_T$\label{Fig. Z_T}]{
%\begin{minipage}[tpb]{5cm}
\begin{minipage}[b]{1\linewidth}
\centering
\includegraphics[scale=0.4]{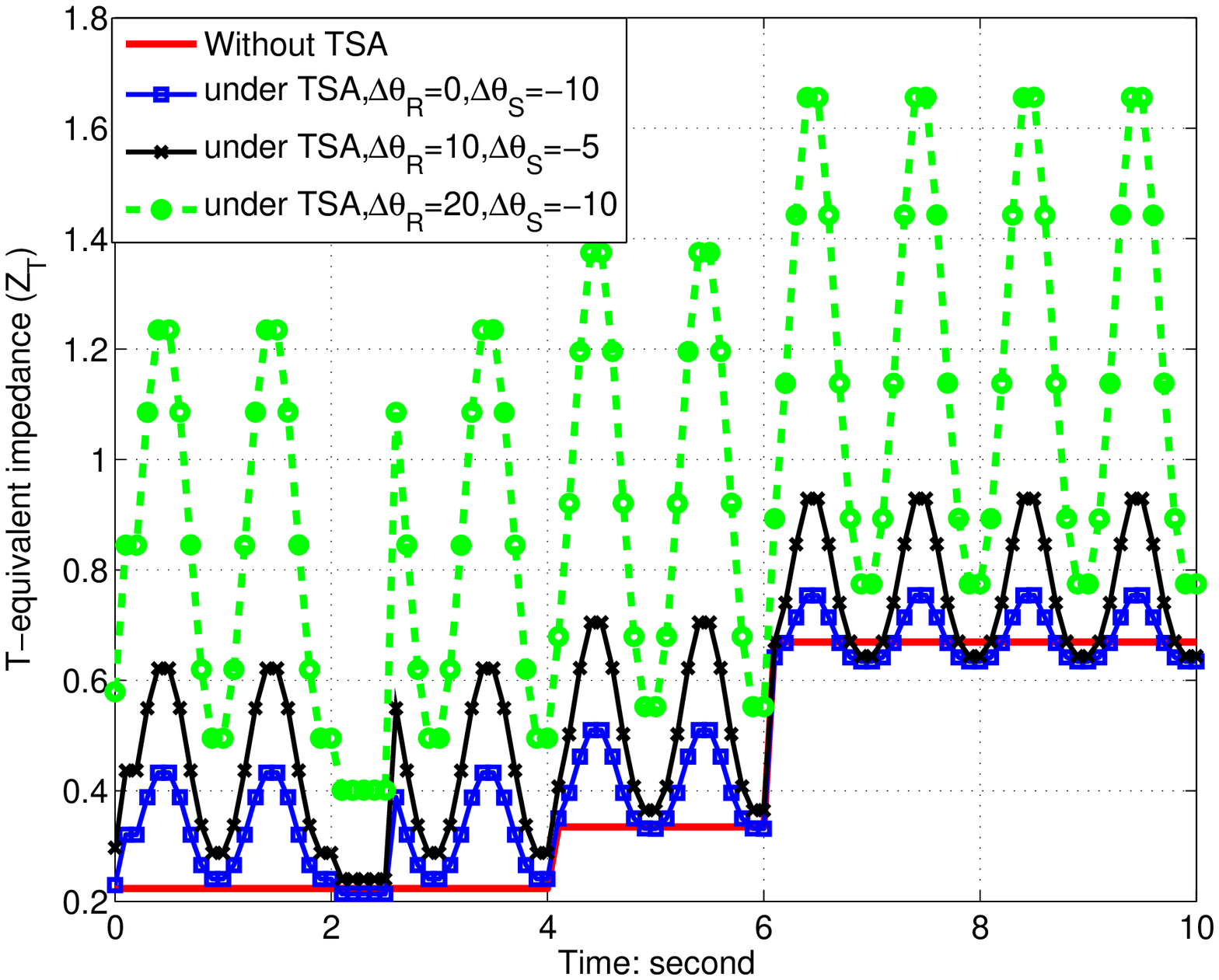}
\end{minipage}}\\%
\subfigure[TSA impact on $Z_{sh}$\label{Fig. Z_sh}]{
%\begin{minipage}[tpb]{11cm}
\begin{minipage}[b]{1\linewidth}
\centering
\includegraphics[scale=0.38]{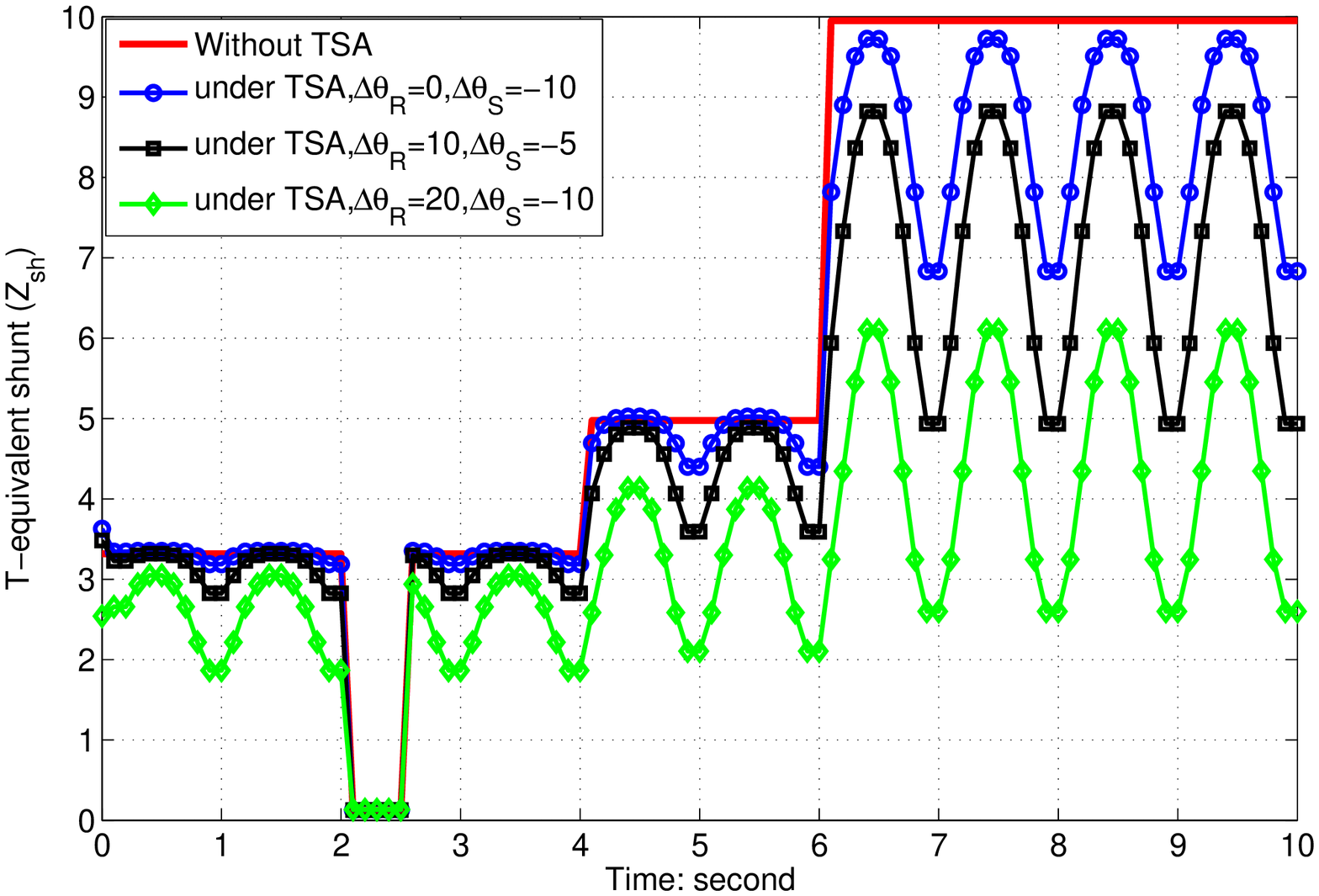}
\end{minipage}}
\caption{Impacts of TSA on the parameters calculation in T-equivalent circuit with different attack strategies} \label{Fig.TSA_Teq}
\end{figure}

It should be noted that the voltage stability indicators are calculated based on $Z_T$ and $Z_{sh}$. Figure \ref{Fig.TSA_Teq} shows the impacts of TSA on the calculation of the
T-equivalent circuit parameters $Z_T$ and $Z_{sh}$.
Without TSA, there are two sharp steps in $Z_T$, which are due to the line trippings.
However, TSA makes these obvious line tripping symptoms ambiguous.
The impact of TSA on the T-equivalent circuit parameters can be considered as having amplitude modulations upon $Z_T$ and $Z_{sh}$.

\begin{figure*}[]
\vspace{0pt}
\subfigure[TSA impact on $Z_{th}$\label{Fig. Z_th}]{
%\begin{minipage}[tpb]{5cm}
\begin{minipage}[b]{0.3\linewidth}
\centering
\includegraphics[scale=0.4]{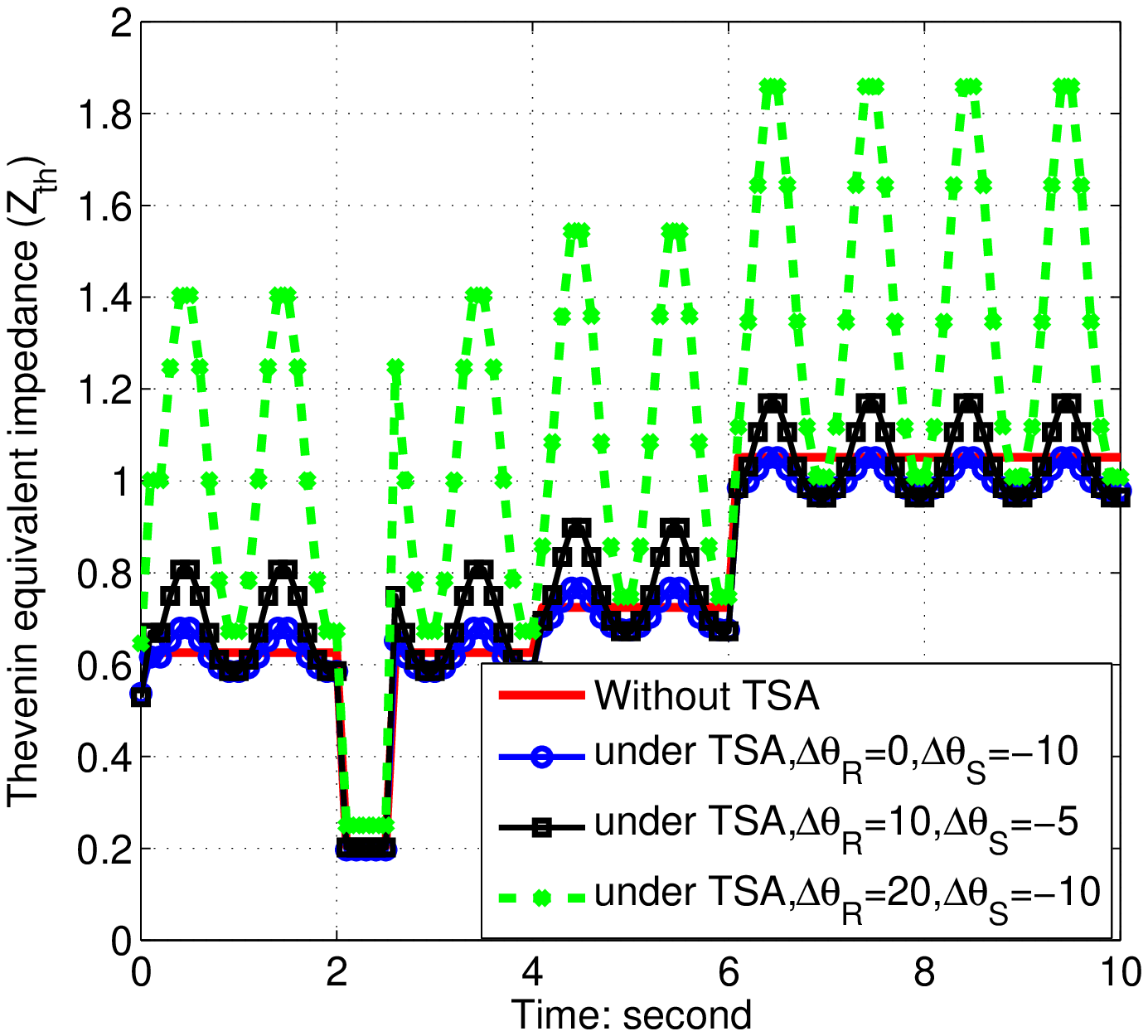}
\end{minipage}}%
%\hfill \vspace{0pt}
\hspace{0.5cm}
\subfigure[TSA impact on the amplitude of $E_{th}$\label{Fig. E_th}]{
%\begin{minipage}[tpb]{11cm}
\begin{minipage}[b]{0.3\linewidth}
\centering
\includegraphics[scale=0.4]{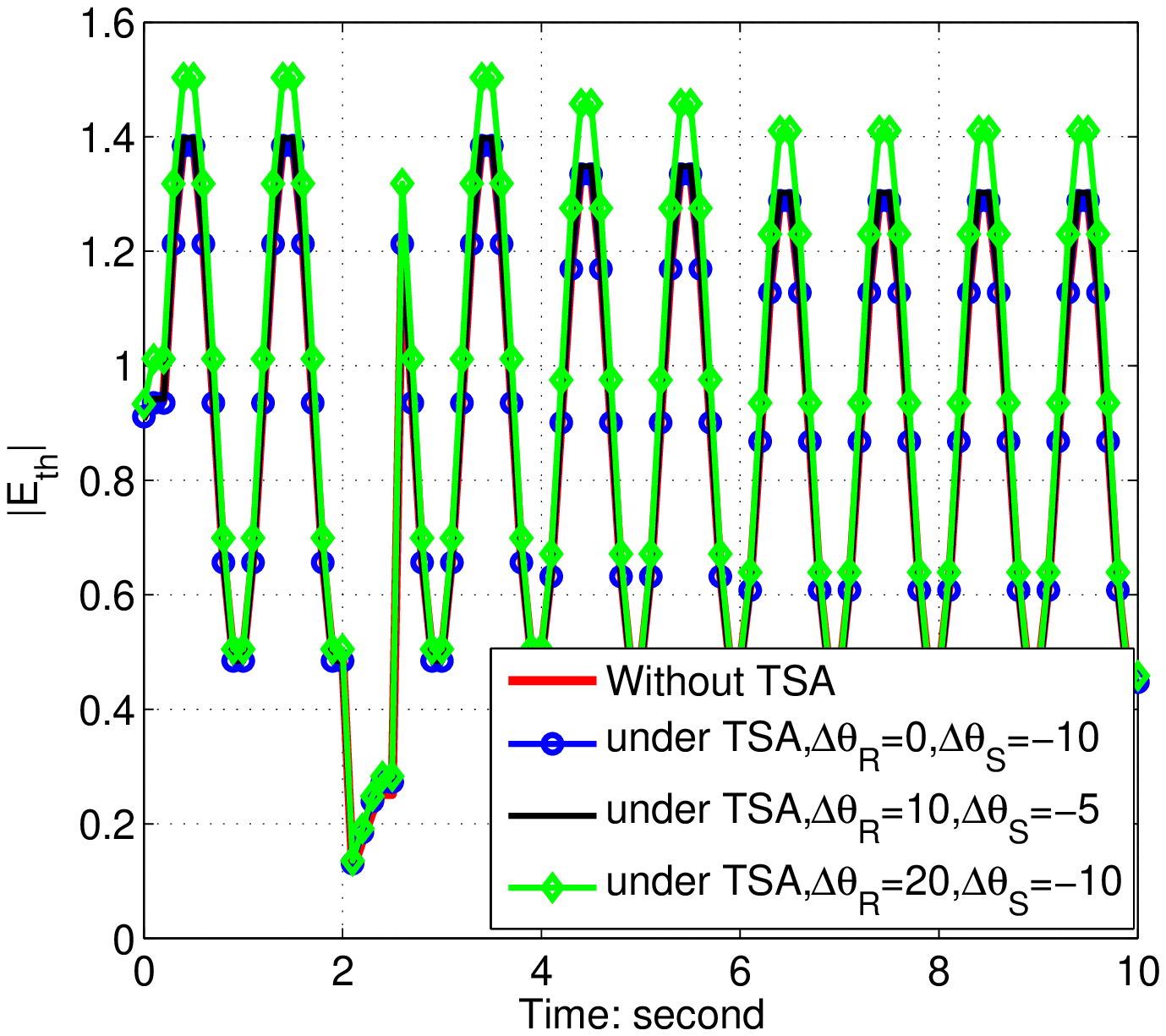}
\end{minipage}}
\hspace{0.5cm}
\subfigure[TSA impact on the phase of $E_{th}$\label{Fig. E_th}]{
%\begin{minipage}[tpb]{11cm}
\begin{minipage}[b]{0.3\linewidth}
\centering
\includegraphics[scale=0.4]{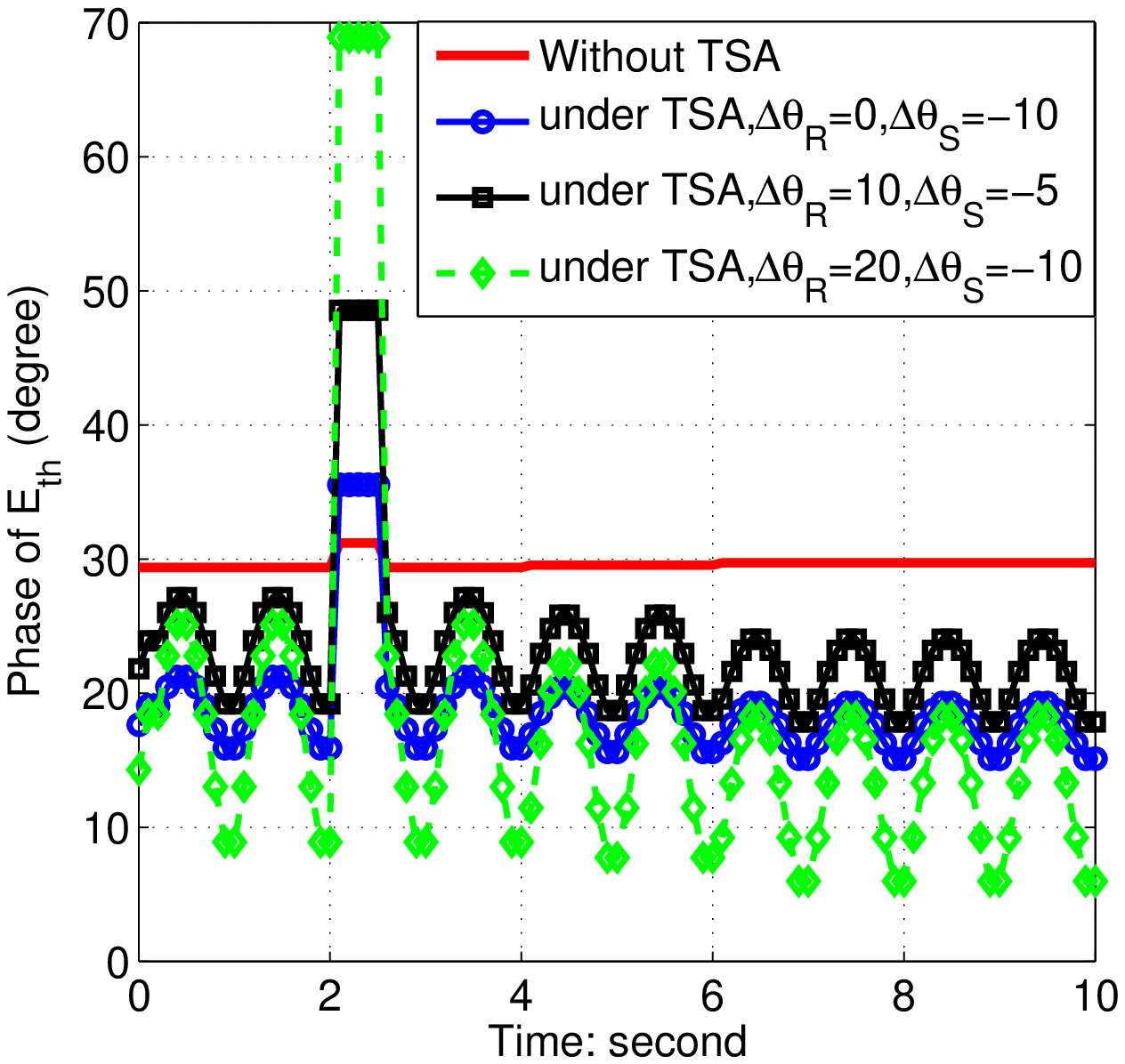}
\end{minipage}}
\caption{Impacts of TSA on the parameters calculation in Thevenin equivalent circuit with different attack strategy} \label{Fig.TSA_TH}
\end{figure*}

The further impact of TSA on the Thevenin equivalent circuit parameters calculation is
shown in Figure \ref{Fig.TSA_TH}. It can be observed from Figure \ref{Fig.TSA_TH} that
TSA has a significant impact on the Thevenin equivalent impedance $Z_{th}$ and the
phase of the Thevenin equivalent voltage source $E_{th}$. The impacts of TSA are similar to those in the T-equivalent circuit, which have amplitude modulations on the parameters.

\begin{figure}[]
\vspace{0pt}
\subfigure[TSA impact on load impedance margin $M_z$\label{Fig. Mz}]{
%\begin{minipage}[tpb]{5cm}
\begin{minipage}[b]{1\linewidth}
\centering
\includegraphics[scale=0.34]{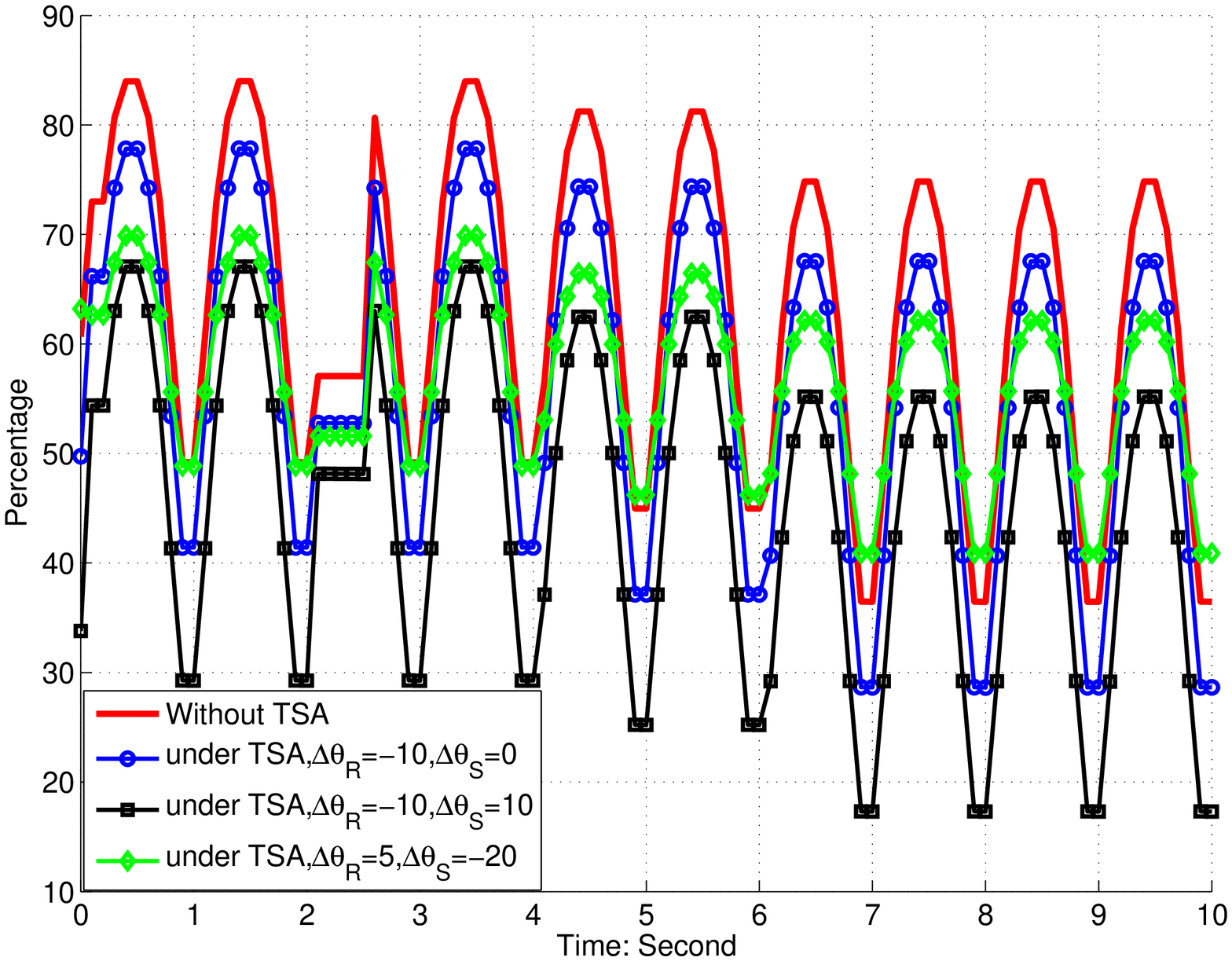}
\end{minipage}}\\%
\subfigure[TSA impact on active power delivered margin $M_p$\label{Fig. Mp}]{
%\begin{minipage}[tpb]{11cm}
\begin{minipage}[b]{1\linewidth}
\centering
\includegraphics[scale=0.34]{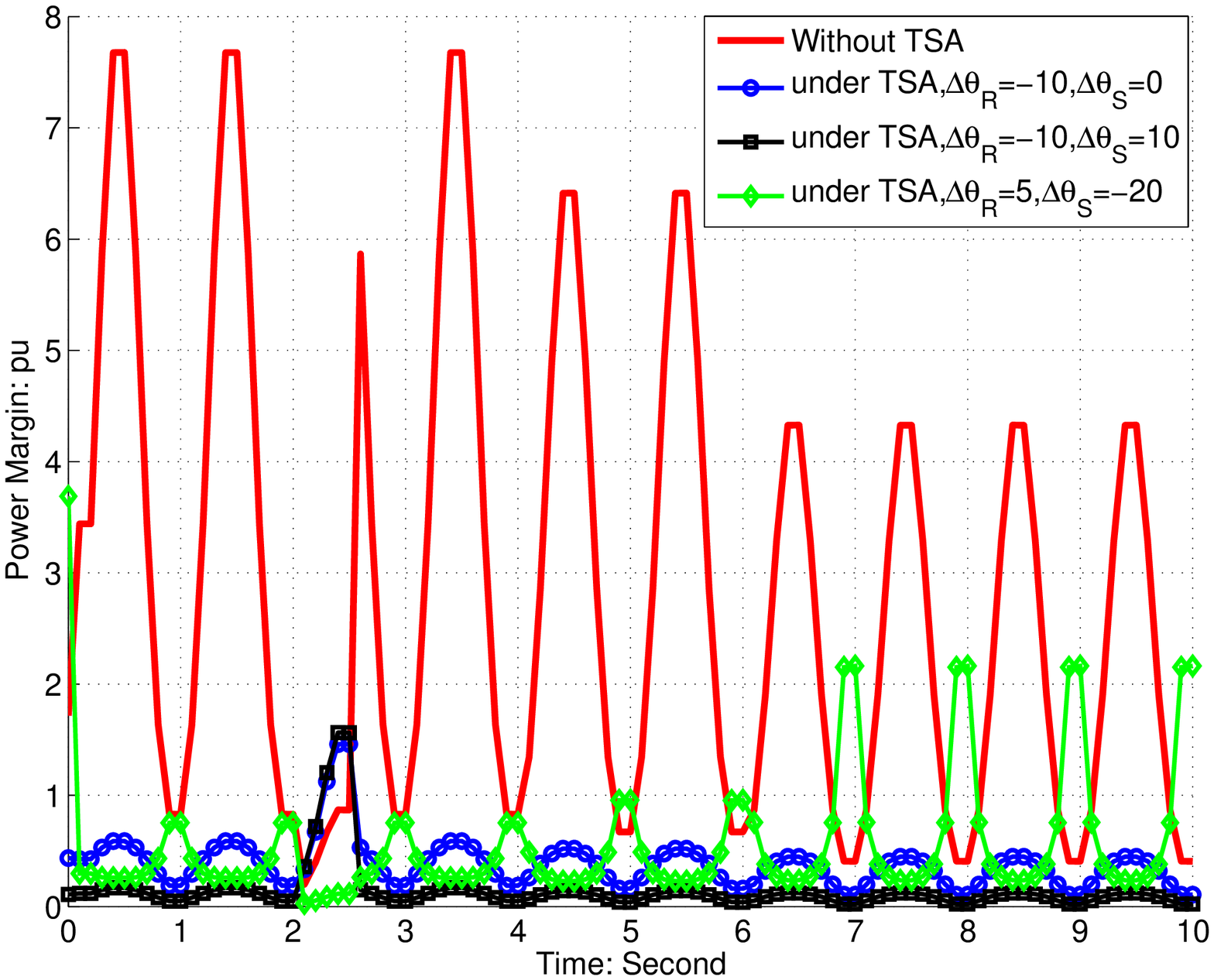}
\end{minipage}}
\caption{TSA impacts on the voltage stability indicators} \label{Fig.TSA_Margin}
\end{figure}

The impacts of TSA on voltage stability indicators are demonstrated
in Fig. \ref{Fig.TSA_Margin} with different attack strategies.
It can be observed that the margin of active delivered power has been greatly
reduced due to the TSA, which misleads the system to implement wrong actions of
voltage stabilization.

\section{TSA in Regional Disturbing Event Location}\label{sec:Event}
In this section, we identify the impact of TSA on regional disturbing event location in smart grid.
One of the important tasks in smart grid is to locate the disturbing event in smart grid in a short time, and consequent isolation will be implemented to prevent cascading failure from spreading to the entire power network.
The disturbing event location is based on the time of arrival (TOA) algorithm \cite{TOA_analysis},
which requires accurate event arrival time. Therefore, TSA has a significant impact on the regional disturbing event location.

\subsection{Principle of Regional Disturbing Event Location}
\begin{figure}[htcp]
  \centering
  \includegraphics[scale=0.5]{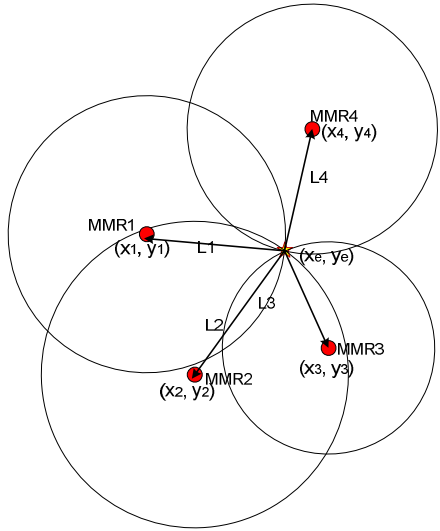}
  \caption{Illustration of regional disturbing event location}\label{Fig:event_ill}
\end{figure}

When a significant disturbance occurs, there will be many symptoms such as voltage and frequency oscillations in both time and space.
The perturbation will travel throughout the grid \cite{V_event}.
Therefore, the distributed monitoring devices can capture the variance of the measurements and send these data to the monitoring system server or exchange with its neighbors. The event time and location can be deduced from the time stamps with these measurements.
After receiving the measurements from these monitoring devices, the servers need to decide the hypocenter of the event, which is typically marked as the wave front arrival time \cite{Event_7}.
By aligning these measurements according to their time stamps, the event arriving time on each monitoring device can be attained. Consequently, the disturbing event location can be deduced by triangulation, which is illustrated in Figure \ref{Fig:event_ill} when there are four PMUs for the event locationing.
The disturbing event location can be derived from solving the following equations when four PMUs are involved
\begin{eqnarray}\label{Eq. location}
\left\{
\begin{array}{llll}
  (x_1-x_e)^2+(y_1-y_e)^2-V_e^2(t_1-t_e)^2 &=& 0 \nonumber\\
  (x_2-x_e)^2+(y_2-y_e)^2-V_e^2(t_2-t_e)^2 &=& 0 \nonumber\\
  (x_3-x_e)^2+(y_3-y_e)^2-V_e^2(t_3-t_e)^2 &=& 0 \nonumber\\
  (x_4-x_e)^2+(y_4-y_e)^2-V_e^2(t_4-t_e)^2 &=& 0,
  \end{array}
  \right.,
\end{eqnarray}
when $t_i, i=1,2,3,4$ is the disturbing event arrival time to the $i$-th PMU, $(x_i,y_i)$ and $(x_e,y_e)$ are the coordinates of the $i$-th PMU and the disturbing event location, respectively; $V_e$ is the event propagation speed in the power grid network. Since the coordinates and the arrival time of each PMU are known, Newtion's method can be applied to solve these equations to attain the event location and time.
Since the sampling is trigged by the GPS receiving signal, a forged GPS time signal can control the sampling in a wrong time and provide wrong time stamps for the measurements.

\subsection{Analysis of Impact}
The principle to obtain the event location coordination and the event time is the TOA algorithm. Since the event monitoring devices in power network are allocated far away from each other, it is difficult to launch cooperative TSA. In this paper,
we analyze the scenario of a single TSA attacker to the system.
We assume that PMU-1 is suffering form TSA, and the arrival time of PMU-1 is modified as
\begin{equation}\label{Eq.t_1}
    t_1=t_1^0+\Delta t,
\end{equation}
where $t_1^0$ is the true arrival time of PMU-1, and $\Delta t$ is the time error due to the TSA. We set $(x_1,y_1)$ as the origin of the transform coordinate for simplicity of analysis \cite{TOA_analysis}. We also set $(x_2,y_2)$ and $(x_3,y_3)$ as $(a,0)$ and $(b,c)$ in the transform coordinates, respectively,
where $a=\sqrt{(x_1-x_2)^2+(y_1-y_2)^2}$, and $b$ and $c$ can be easily
changed into the new coordinates by using the follow equations:
\begin{eqnarray}\label{Eq.Transform_Coordiant}
  b &=& (x_3-x_1)\cos\alpha+(y_3-y_1)\sin\alpha \\
  c &=& -(x_3-x_1)\sin\alpha+(y_3-y_1)\cos\alpha,
\end{eqnarray}
where
\begin{equation}\label{Eq.Alpha}
    \alpha = \tan^{-1}\left(\frac{y_2-y_1}{x_2-x_1}\right).
\end{equation}
We define $k^2=x_e^{'2}+y_e^{'2}$, where $(x'_e,y'_e)$ is the transformed coordinate for the event location. Similarly to the analysis in \cite{TOA_analysis}, we define two pseudo-ranges
$L=(t_2-t_1)V_e$ and $R=(t_3-t_1)V_e$. It is easy to obtain the close form of the solution, which is given by
\begin{eqnarray}\label{Eq.E_xy}
  x'_e &=& A+Bk \\
  y'_e &=& C+Dk,
\end{eqnarray}
where
\begin{eqnarray}
  A &=& \frac{a^2-L^2}{2a} \\
  B &=& -\frac{L}{a} \\
  C &=& \frac{b^2+c^2-2bA-R^2}{2c} \\
  D &=& -\frac{R+bB}{c}.
\end{eqnarray}
It is easy to transform the coordinate of the event location into the original coordinate, which is given by
\begin{eqnarray}
  x_e &=& x'_e\cos\alpha-y'_e\sin\alpha+x_1 \\
  y_e &=& x'_e\sin\alpha-y'_e\cos\alpha+y_1.
\end{eqnarray}
Since TSA only affects PMU-1, we analyze how $t_1$ affects the location error.
The partial derivatives $x'_e$  and $y'_e$ with respect to $t_1$ are given by
(\ref{Eq.dx}) and (\ref{Eq.dy}).

\newcounter{mytempeqncnt}
\begin{figure*}[!t]
% ensure that we have normalsize text
\normalsize
% Store the current equation number.
\setcounter{mytempeqncnt}{\value{equation}}
% Set the equation number to one less than the one
% desired for the first equation here.
% The value here will have to changed if equations
% are added or removed prior to the place these
% equations are referenced in the main text.
\setcounter{equation}{55}
\begin{eqnarray}\label{Eq.dx}
   \delta x(t1) &=& \frac{\partial{x_e}}{\partial{t_1}}\nonumber\\
   &=& \frac{L}{a}V_e+\frac{K}{a}V_e \nonumber\\
   &&  +BM\frac{-(AV_e/a-BLV_e/a+C/c(V_e-bV_e/a)+DV_e(-bL/ac+R/c))}            {(B^2+D^2-1)^2}\nonumber\\
   &&  \mp BM\frac{4NV_e(A/a+LB/a+C/c(1-b/a)+D(-bL/(ac)+R/c))}
      {4(B^2+D^2-1)^2\sqrt{N^2-4MP}}\nonumber\\
   &&  \pm 8B\frac{MV_e(AL/a+C(-bL/ac+R/c))+PV_e(B/a+D/c(1-b/a))}
      {4(B^2+D^2-1)^2\sqrt{N^2-4MP}}\nonumber\\
   && +BV_e(B/a+D/c(1-b/a))(N\pm\sqrt{N^2-4MP}),
\end{eqnarray}
\begin{eqnarray}\label{Eq.dy}
 \delta y(t1) &=& \frac{\partial{y_e}}{\partial{t_1}}\nonumber\\
   &=& \frac{V_e}{c}(R-bL/a+(1-b/a)k)\nonumber\\
   &&  +DM\frac{-(AV_e/a-BLV_e/a+C/c(V_e-bV_e/a)+DV_e(-bL/ac+R/c))}{(B^2+D^2-1)^2}\nonumber\\
   &&  \mp DM\frac{4NV_e(A/a+LB/a+C/c(1-b/a)+D(-bL/(ac)+R/c))}
      {4(B^2+D^2-1)^2\sqrt{N^2-4MP}}\nonumber\\
   &&  \pm 8D\frac{MV_e(AL/a+C(-bL/ac+R/c))+PV_e(B/a+D/c(1-b/a))}
      {4(B^2+D^2-1)^2\sqrt{N^2-4MP}}\nonumber\\
   && +DV_e(B/a+D/c(1-b/a))(N\pm\sqrt{N^2-4MP}),
\end{eqnarray}

% Restore the current equation number.
\setcounter{equation}{57}
% IEEE uses as a separator
\hrulefill
% The spacer can be tweaked to stop underfull vboxes.
\vspace*{4pt}
\end{figure*}
The parameter $N$, $M$, and $P$ can further expressed as:
\begin{eqnarray}
  N &=& AB+CD \\
  M &=& B^2+D^2-1\\
  P &=& A^2+C^2.
\end{eqnarray}
After obtaining the partial differentiation in the transform coordinate,
it is easy to obtain the partial differentiations in the original coordinate,
which are given by
\begin{eqnarray}
  \frac{\partial{x_e}}{\partial{t_1}} &=& \delta x(t1)\cos\alpha-\delta y(t1)\sin\alpha \\
  \frac{\partial{y_e}}{\partial{t_1}} &=& \delta x(t1)\sin\alpha+\delta y(t1)\cos\alpha
\end{eqnarray}

\subsection{Simulation Results}
\begin{figure}[htcp]
  \centering
  \includegraphics[scale=0.3]{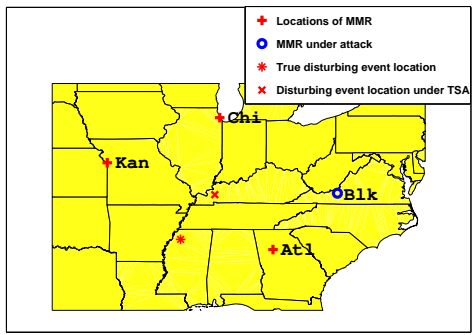}
  \caption{Simulation of TSA on disturbing event location}\label{Fig:Event_map}
\end{figure}

For the disturbing event location, the sampling is trigged by the GPS time signal as illustrated in Figure \ref{Fig:SynMeasure}. A forged GPS time signal can control the sampling in a wrong time or provide a wrong time stamp for the measurements. The simulation illustrating the impact on the event location is shown in Figure \ref{Fig:Event_map}. It is observed that, with one PMU under TSA, the estimation of disturbing event will be far away from the true position (the event happening in Mississippi is misled to Tennessee).

\begin{figure}[htcp]
  \centering
  \includegraphics[scale=0.4]{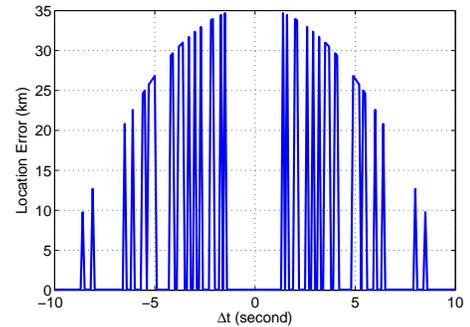}
  \caption{Location error under various $\Delta t$}\label{Fig:Event_Error}
\end{figure}

Based on the analytical results, we simulate the location error with different
$\Delta t$, which is given by Figure \ref{Fig:Event_Error}. It is observed that
the location error caused by TSA is nonlinear.

\section{Conclusion}\label{sec:Conclusion}
In this paper, we have identified the GPS spoofing based TSA in power grids. The time stamps are modified by the forged GPS signal, and the measurements with time stamps will be corrupted by TSA. For several scenarios, the impacts of TSA have been studied. For the transmission line fault detection and location, TSA can not only deteriorate the performance of fault location, but also increase the false alarm probability with some fault indicators. For the voltage stability monitoring, TSA can exaggerate the power margin and result in delaying or disabling the voltage instability alarm. It has also been demonstrated that the TSA can significantly damage the event location in power grid.

%%%%%%%%%%%%%%%%%%%%% Refence list %%%%%%%%%%%%%%%%%%%%%%%%%%%%%%%%%%%%%%%%%%

%%%%%%%%%%%%%%% Set of figures

%%%%%%%%%%%%%%%%%%%%%%%%%%%%%%%%%%%%%%%%%%%%%%%%%%%%%%%%%%%%%%%%%%%%%%%%%%%%%%%%%
%% biography section
%%
%% If you have an EPS/PDF photo (graphicx package needed) extra braces are
%% needed around the contents of the optional argument to biography to prevent
%% the LaTeX parser from getting confused when it sees the complicated
%% \includegraphics command within an optional argument. (You could create
%% your own custom macro containing the \includegraphics command to make things
%% simpler here.)
%%\begin{biography}[{\includegraphics[width=1in,height=1.25in,clip,keepaspectratio]{mshell}}]{Michael Shell}
%% or if you just want to reserve a space for a photo:
%\end{IEEEbiography}
%% insert where needed to balance the two columns on the last page with
%%\newpage
%% You can push biographies down or up by placing
%% a \vfill before or after them. The appropriate
%% use of \vfill depends on what kind of text is
%% on the last page and whether or not the columns
%% are being equalized.
%
%%\vfill
%
%% Can be used to pull up biographies so that the bottom of the last one
%% is flush with the other column.
%%\enlargethispage{-5in}
%
%
%
%% that's all folks
\end{document}